\documentclass{2023SCGE}

\usepackage{graphicx}	
\usepackage{amsmath}	
\usepackage[table,xcdraw,dvipsnames]{xcolor}
\usepackage{hyperref}
\usepackage{float}
\usepackage{multirow}

\DeclareGraphicsExtensions{.pdf,.png,.jpg}

\newcommand{\HI}{\text{H\,\textsc{i}} }

\DeclareRobustCommand{\ion}[2]{%
\relax\ifmmode
\ifx\testbx\f@series
{\mathbf{#1\,\mathsc{#2}}}\else
{\mathrm{#1\,\mathsc{#2}}}\fi
\else\textup{#1\,{\mdseries\textsc{#2}}}%
\fi}

\begin{document}

\ensubject{subject}

\ArticleType{Article}
\Year{2026}
\Month{??}
\Vol{??}
\No{??}
\DOI{??}
\ArtNo{000000}
\ReceiveDate{??}
\AcceptDate{??}

\title{\boldmath No way ou$\tau$: Epoch of Reionization Observations Do not Support\\ Large Values of the Optical Depth to Reionization}

\author[1]{\\Paulo Montero-Camacho\thanks{Email: \href{mailto:pmontero@pcl.ac.cn}{pmontero@pcl.ac.cn}}}{}%
\author[2]{Yi Mao\thanks{Email: \href{mailto:ymao@tsinghua.edu.cn}{ymao@tsinghua.edu.cn}}}{}

\AuthorMark{Montero-Camacho}

\AuthorCitation{Montero-Camacho, et al}

\address[1]{Department of Strategic and Advanced Interdisciplinary Research, \\ Pengcheng Laboratory, Nanshan District, Shenzhen, Guangdong 518000, China}
\address[2]{Department of Astronomy, Tsinghua University, Beijing 100084, China}

\abstract{
    Recent cosmological analyses combining high-redshift cosmic microwave background (CMB) measurements with low-redshift baryon acoustic oscillation (BAO) data have reported a preference for dynamical dark energy, with the cosmological constant scenario ($\Lambda$CDM) disfavored at the $\sim 3\sigma$ level. These analyses, however, typically rely on large-scale CMB polarization measurements to constrain the optical depth to reionization $\tau_{\rm reio}\sim 0.06$, raising the question of whether potential systematics in this dataset could influence the inferred cosmological preference. Excluding large-scale polarization data substantially weakens the tension with $\Lambda$CDM to the $\lesssim 2\sigma$ level, nevertheless at a price of increasing $\tau_{\rm reio}$ significantly to $\sim 0.09$. Here, we use a physically motivated Gompertzian reionization framework to perform a self-consistent Bayesian analysis combining CMB (excluding large-scale polarization data), BAO, and independent measurements of the neutral hydrogen fraction evolution from quasar damping wing observations and dark pixel constraints. We derive $\tau_{\rm reio} = 0.067 \pm 0.011$ (dynamical dark energy scenario) in good agreement with cosmological analyses that would include large-scale CMB polarization data, while the inferred reionization history is consistent with multiple observational constraints. Our analysis recovers a preference for dynamical dark energy at the $\gtrapprox2\sigma$ level. These results demonstrate that astrophysical probes of reionization independently recover the optical depth required by CMB polarization measurements, suggesting that potential systematics in large-scale polarization alone are unlikely to fully explain the emerging preference for dynamical dark energy.
}

\vspace{1cm}

\vspace{0.2cm}

\maketitle

\tableofcontents

\begin{multicols}{2}

\section{Introduction}
\label{sec:intro}
The standard cosmological model ($\Lambda$CDM) with its cosmological constant has been remarkably successful in describing the evolution of the Universe and a wide range of cosmological observables, such as the Cosmic Microwave Background (CMB). However, recent measurements from the Dark Energy Spectroscopic Instrument (DESI; \cite{2022AJ....164..207D}) indicate a preference for dynamical dark energy ($w_0w_a$CDM), leading to an emerging $\sim 3\sigma$ tension with $\Lambda$CDM \cite{2025PhRvD.112h3515A}\footnote{The inclusion of full-shape DESI DR1 data reduces this tension to $1.4\sigma$ and $1\sigma$ for power spectrum \cite{2026arXiv260218761F} and bispectrum \cite{2026arXiv260623936F}, respectively.}. 

In this work, we investigate the exclusion of large-scale CMB polarization data as a possible mechanism for resolving this tension \cite{2026PhRvL.136h1002S,2025PhRvD.112d3541J,2026PhRvD.113h3515A,2025arXiv250821069E,2026arXiv260109644K}. Excluding these data allows the optical depth to reionization $\tau_{\rm reio}$ to increase to $\sim 0.09$. Since the CMB-inferred $\tau_{\rm reio}$ is negatively correlated with the total matter density, $\Omega_{\rm m}$, a higher value of $\tau_{\rm reio}$ can accommodate a lower $\Omega_{\rm m}$, thus reducing the discrepancy between early and late-time measurements of $\Omega_{\rm m}$ \cite{2025ApJ...983L..27T} and potentially alleviating the $H_0$ tension simultaneously \cite{2025JCAP...08..082A}. However, large-scale CMB polarization data, last measured by the Planck collaboration \cite{2020A&A...641A...5P} and routinely included in modern CMB analyses, tightly constrain $\tau_{\rm reio}$ to $\sim 0.06$ \cite{2020A&A...635A..99P,2025JCAP...11..062L}\footnote{The preference for $\tau_{\rm reio} \sim 0.06$ remains even when extending the parameter space, e.g., by allowing for running of the spectral index, varying the effective number of non-photon radiation species, and considering quintessence-like dark energy models (see, e.g., \cite{2024ApJ...976L..11R,2025ApJ...986L..31R,2025ApJ...994L..26R}).}, hence preventing the exploitation of this degeneracy. Furthermore, when the interplay between reionization and cosmological parameters is modeled self-consistently, the $\tau_{\rm reio} - \Omega_{\rm m}$ degeneracy becomes significantly weaker (see Figure 1 of \cite{2026arXiv260413423M}).

Motivated by the large value of $\tau_{\rm reio}$ reported by \cite{2026PhRvL.136h1002S,2025PhRvD.112d3541J}, several mechanisms have been proposed or revisited to explain such an increase. For instance, Ref.~\cite{2026arXiv260620795J} introduces inflationary features that enhance the value of $\tau_{\rm reio}$, Ref.~\cite{2026JCAP...06..019U} considers the impact of infrared cutoff models, and Ref.~\cite{2026arXiv260619459A} suggests that supermassive Population III stars could produce an early contribution to reionization while remaining consistent with constraints from the end stages of reionization. 

Refs. \cite{2026PhRvL.136h1002S,2025PhRvD.112d3541J} showed that excluding large-scale CMB polarization data relaxes the $3\sigma$ tension with $\Lambda$CDM to $\lesssim2\sigma$. At the same time, it alleviates the apparent preference for negative neutrino masses and weakens the exclusion of the $\sum m_\nu \geq 0.06$ eV lower bound implied by neutrino oscillation experiments, which is otherwise disfavored by approximately $\sim3\sigma$ \cite{2025PhRvD.112h3513E,2025PhRvD.111h3507G,2026arXiv260413423M}. Measuring CMB polarization is considerably challenging as the signal is several orders of magnitude weaker than temperature anisotropies, requiring careful treatment of instrumental systematics and astrophysical foregrounds \cite{2019A&A...629A..38D,2020A&A...641A...1P,2020A&A...635A..99P,2021A&A...651A..65L}. Nevertheless, no evidence of unresolved systematic effects in the low-$\ell$ EE power spectrum has been reported to date.

Among the primary free parameters of the standard cosmological model, $\tau_{\rm reio}$ remains the only one that has not yet been measured with percent-level precision (see Figure 1 in \cite{2024arXiv240513680M} for recent developments). It quantifies the cumulative Thomson scattering experienced by CMB photons as they propagate towards our telescopes. As such, $\tau_{\rm reio}$ is not a fundamental cosmological parameter but rather an effective quantity determined by the ionization history of the Universe. Consequently, it can technically be demoted to derived parameter rather than a free parameter \cite{2024arXiv240513680M,2025PhRvD.112d3506F,2026arXiv260413423M}. Its inferred value depends on the assumed reionization history $x_\HI(z)$.

Most CMB analyses model the reionization history using a hyperbolic tangent, a fast symmetric sigmoid that does not reflect the behavior predicted by reionization simulations \cite{2011MNRAS.411..955M,2022MNRAS.511.4005K,2023MNRAS.519.6162P,2026arXiv260515310Z}. Simulations instead favor an asymmetric evolution, with a gradual start driven by the scarcity of ionizing sources at high redshifts. Furthermore, the $\tanh$ prescription performs poorly in joint analysis combining CMB and epoch of reionization (EoR) observations (see Appendix C of \cite{2026arXiv260413423M} and Figure \ref{fig:xHI}). When only CMB data are considered, the effect of adopting an asymmetric reionization history can be modest, altering the large-scale polarization signal by $\lesssim 4\%$ \cite{2016A&A...596A.108P}. Notably, early asymmetric parameterizations relied on generic functional forms chosen to reproduce the expected asymmetry \cite{2015A&A...580L...4D,2018ApJ...858L..11T}. More recent models, however, are calibrated directly against reionization simulations, such as the Gompertzian reionization model introduced in \cite{2024arXiv240513680M}. Relative to the $\tanh$ prescription, the Gompertzian model provides a substantially better description of joint CMB + EoR data and typically yields lower values of $\tau_{\rm reio}$.

As an alternative to simulation-informed reionization models, such as the Gompertzian prescription or neutral network-based approaches \cite{2025PhRvD.112d3506F},  Refs.~\cite{2025arXiv250821069E,2026arXiv260109644K} reconstructed the reionization history directly from publicly available $x_\HI$ measurements. This strategy implicitly treats all $x_\HI$ observations as equally reliable, although their robustness can vary significantly (e.g., \cite{2023ApJ...953...29B}). Moreover, these generic reconstructions do not incorporate physical insights from our current understanding of the reionization process. Nevertheless, both studies found that including a broad compilation of $x_\HI$ constraints favors low values of the optical depth, thus preserving the $\Lambda$CDM tension.

Even if some issue were identified in the large-scale CMB polarization data, current EoR observations appear broadly consistent with the optical depths inferred from those measurements \cite{2026arXiv260109644K}. Consequently, any systematic affecting the low-$\ell$ CMB EE power spectrum would also need to be accompanied by inconsistencies in multiple independent EoR observations. In this work, we investigate whether including robust EoR constraints -- while still excluding large-scale CMB polarization data -- and adopting the physically motivated Gompertzian reionization model alters the reported relaxation of the tension with $\Lambda$CDM. 

The remainder of this paper is organized as follows. We describe the data and methodology used throughout this work in Section \ref{sec:data}. In Section \ref{sec:tau_lcdm}, we present a comparison of the optical depth inferences obtained for different reionization models within $\Lambda$CDM. Section \ref{sec:w0wa} reports our main results, i.e., the effect of a joint analysis of CMB and robust epoch of reionization data on dynamical dark energy preferences within a physically motivated Gompertzian reionization model. We summarize our findings in Section \ref{sec:conclusion}. \ref{app:gomp} introduces the Gompertzian reionization model established in \cite{2024arXiv240513680M}.

\begin{table*}[!htb]
\centering
\resizebox{17.5cm}{!}{

    \centering
    \begin{tabular}{l|cccccc}
    \toprule
    {} & $\Lambda$T & $\Lambda$G* & $\Lambda$G & $w_0w_a$T & $w_0w_a$G & DESI DR2  \\
    \midrule
    Model & $\Lambda$CDM & $\Lambda$CDM & $\Lambda$CDM & $w_0w_a$CDM & $w_0w_a$CDM & $w_0w_a$CDM \\
    Reionization & $\tanh$ & Gompertzian & Gompertzian & $\tanh$ & Gompertzian & $\tanh$ \\
    Data & $\not\ni$ low$\ell$ $C_\ell^{\rm EE}$ & $\not\ni$ low$\ell$ $C_\ell^{\rm EE}$ & $\not\ni$ low$\ell$ $C_\ell^{\rm EE}$, $\ni$ DW + DP & $\not\ni$ low$\ell$ $C_\ell^{\rm EE}$ & $\not\ni$ low$\ell$ $C_\ell^{\rm EE}$, $\ni$ DW + DP & $\ni$ low$\ell$ $C_\ell^{\rm EE}$ \\
    $\tau_{\rm reio}$ & $0.090 \pm 0.012$ & $0.087 \pm 0.011$ & $0.077 \pm 0.011$ & $0.0870 \pm 0.0056$ & $0.067 \pm 0.011$ & $\sim 0.06$\\
    Informative $\tau_{\rm reio}$ prior & $\mathcal{U}(0.02, 0.2)$ & Derived param. & Derived param. & $\mathcal{N}(0.09,0.006)$ & Derived param. & $\mathcal{U}(0.01, 0.8)$ \\
    $\Lambda$CDM tension & & & & $<2\sigma$ & $\gtrapprox2\sigma$ & $>3\sigma$  \\
    Reference & \cite{2026PhRvL.136h1002S} & & & \cite{2026PhRvL.136h1002S} & & \cite{2025PhRvD.112h3515A}\\
    \bottomrule
    \end{tabular}}
    \caption{Models considered throughout this work in the context of both  results from DESI DR2 and Ref.~\cite{2026PhRvL.136h1002S}. The former established the current $w_0w_a$CDM$-\Lambda$CDM tension, while the latter proposes a potential resolution through the exclusion of large-scale CMB polarization data. Note that our work does not impose any priors on the optical depth to reionization since it is a derived parameter within the Gompertzian reionization framework.}
    \label{tab:sum}
\end{table*}

\section{Data and Method}
\label{sec:data}
For BAO, we use the DESI DR2 measurements \cite{2025PhRvD.112h3515A}. For the high-$\ell$ CMB TT, TE, and EE likelihood, we adopt the combination of Planck PR4 \cite{2019arXiv191000483E} and ACT DR6 \cite{2025JCAP...11..062L}, following \cite{2025JCAP...11..062L}. We additionally include the Planck PR3 large-scale ($2 \leq \ell \leq 30$) temperature anisotropy measurements \cite{2020A&A...641A...5P}, which we refer to as low-$\ell$ CMB TT. For CMB lensing, we combine the Planck PR4 \cite{2022JCAP...09..039C} and ACT DR6 \cite{2024ApJ...962..112Q} data, following \cite{2024ApJ...962..112Q}. Moreover, we supplement these cosmological datasets with robust constraints on the reionization history from quasar damping wings (QSO DW; \cite{2022MNRAS.512.5390G,2024MNRAS.530.3208G,2024A&A...688L..26S,2024ApJ...969..162D}) and the two lowest-redshift Lyman-$\alpha$ and Lyman-$\beta$ dark pixel measurements from \cite{2026MNRAS.545f1862D}\footnote{As shown in Figure \ref{fig:xHI_LCDM}, the two lowest redshift dark pixel (DP) bounds provide the strongest model independent constraints on the late stages of reionization.}. 

We employ a physically motivated Gompertzian reionization model \cite{2024arXiv240513680M} to reconstruct the global neutral hydrogen fraction $x_\HI(z)$ as a function of redshift. In contrast to the standard $\tanh$ prescription, which models reionization as a symmetric sigmoid with an often fixed duration and thus a single free parameter corresponding to the midpoint of reionization, the Gompertzian model describes reionization with an asymmetric sigmoid characterized by two free parameters. Its main advantages compared to the hyperbolic tangent prescriptions are that the model is physically motivated as it agrees well with the expectations from EoR simulations and it is well-suited for joint analyses of CMB and EoR observations. The self-consistent inclusion of robust EoR measurements improves constraints on  $\tau_{\rm reio}$, thereby reducing degeneracies with cosmological parameters. We recap the Gompertzian reionization model in \ref{app:gomp}. 

The Gompertzian reionization model has been shown to accurately reproduce the reionization history of simulations spanning a range of astrophysical prescriptions \cite{2024arXiv240513680M}. These simulations, however, assume $\Lambda$CDM. This approximation is well justified because cosmic reionization occurs deep within the matter-dominated era, where the impact of dark energy should be negligible. Although the exact timing of the transition from matter domination to dark energy domination depends weakly on the underlying reionization model, the effect on the reionization history is expected to be small. In fact, Gompertzian reionization generally favor later (lower redshift) transitions to matter domination  than that inferred using the hyperbolic tangent prescription (see Appendix~5 of \cite{2026arXiv260413423M}).

We perform self-consistent joint Monte Carlo Markov Chain (MCMC) analyses using {\sc cobaya} \cite{2021JCAP...05..057T}, combining high-$\ell$ CMB, low-$\ell$ CMB TT, CMB lensing, BAO, QSO DW, and dark pixel data. The Gompertzian reionization model is implemented directly in the Boltzmann solver {\sc CLASS} \cite{2011JCAP...07..034B}\footnote{\url{https://github.com/paulomontero/class_gomp}}, while the implementation of the EoR likelihood follows \cite{2024arXiv240513680M,2026arXiv260413423M}. We denote the Gompertzian $\Lambda$CDM and $w_0w_a$CDM models by $\Lambda$G and $w_0w_a$G, respectively. The $w_0w_a$ model additionally allow the sum of the neutrino masses to vary. We also introduce a benchmark $\Lambda$CDM Gompertzian reionization model $\Lambda$G* which does not include QSO DW nor dark pixel data. For comparison, we refer to the corresponding hyperbolic tangent models from \cite{2026PhRvL.136h1002S}\footnote{These models are available at \url{https://zenodo.org/records/15298950}. $\Lambda$T corresponds to `lcdm\_mnu=0.06\_tau=free\_cmb-p+cmb-l+bao' while $w_0w_a$T is `w0wa\_mnu=0.06\_tau=0.09\_cmb-p+cmb-l+bao'.} as $\Lambda$T and $w_0w_a$T for $\Lambda$CDM and $w_0w_a$CDM, respectively. Note that as Ref.~\cite{2026PhRvL.136h1002S} used the hyperbolic tangent prescription, they do not include EoR observations, instead they sample over $\tau_{\rm reio}$, while $w_0w_a$T also adds a Gaussian prior centered at $\tau_{\rm reio} = 0.09$ (the $\Lambda$T value), which obscures the direct comparison between the different models. Table~\ref{tab:sum} summarizes these models and contextualizes them in terms of the DESI DR2 results.

\section{Optical depth from joint CMB and EoR observations in $\mathbf{\Lambda}$CDM}
\label{sec:tau_lcdm}
\begin{figure}[H]
    \centering
    \includegraphics[width=0.95\columnwidth]{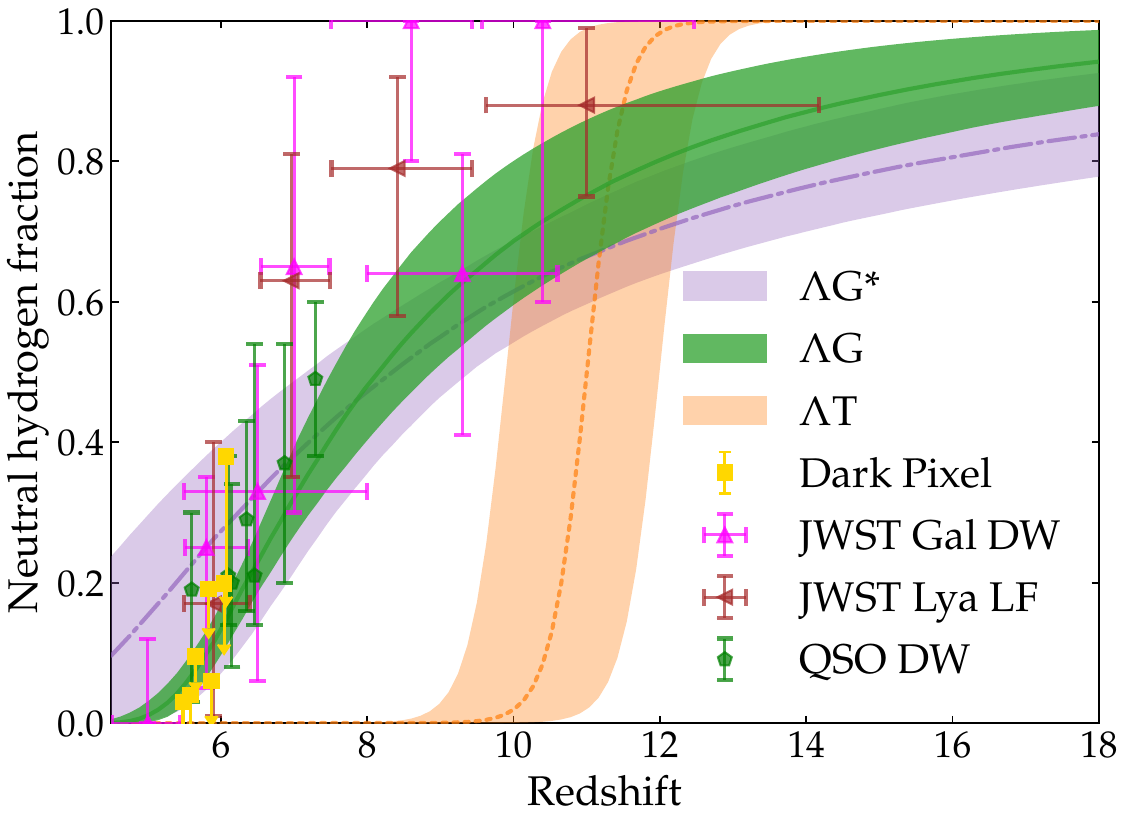}
    \caption{Reionization histories inferred from representative posterior best fits using high-$\ell$ CMB, low-$\ell$ CMB TT, CMB lensing, and BAO as the cosmological dataset. The purple shaded region shows the 1$\sigma$ confidence interval obtained with a Gompertzian reionization model in $\Lambda$CDM without EoR data ($\Lambda$G*). Furthermore, the green shaded region depicts the reionization history obtained from a Gompertzian reionization model in $\Lambda$CDM ($\Lambda$G) with the additional inclusion of  QSO DW and Dark Pixel measurements of the reionization timeline. For comparison, the orange shaded region shows the corresponding $1\sigma$ constraint obtained by Ref.~\cite{2026PhRvL.136h1002S} using a hyperbolic tangent ($\Lambda$T). Representative observational constraints on the epoch of reionization are overlaid, including dark pixel measurements \cite{2015MNRAS.447..499M,2026MNRAS.545f1862D}, quasar and galaxy damping wing observation \cite{2022MNRAS.512.5390G,2024MNRAS.530.3208G,2024A&A...688L..26S,2024ApJ...969..162D,2026A&A...705A.114M,2026ApJ...997...86U}, and Ly$\alpha$ luminosity functions constraints \cite{2021ApJ...919..120M,2025ApJS..278...33K}.}
    \label{fig:xHI_LCDM}
\end{figure}

Figure \ref{fig:xHI_LCDM} shows the inferred reionization history and its corresponding $1\sigma$ confidence level (CL) region in the context of multipole independent EoR observations for both the $\Lambda$T and $\Lambda$G models. To illustrate the constraining power of the EoR data, we also show the $1\sigma$ CL obtained from a $\Lambda$CDM analysis that excludes both large-scale CMB polarization and EoR constraints ($\Lambda$G*). As expected, this analysis yields substantially weaker constraints on $x_\HI$, particularly yielding loose constraints at the end stages of reionization. Consequently, $\tau_{\rm reio}$ shifts to larger values. For comparison, we show the $\Lambda$T result from \cite{2026PhRvL.136h1002S}, which samples over $\tau_{\rm reio}$ directly without the use of any EoR observations. Its derived midpoint of reionization is largely inconsistent with EoR observations. 

\begin{figure}[H]
    \centering
    \includegraphics[width=0.95\columnwidth]{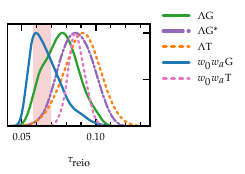}
    \caption{Posterior distributions  of the optical depth to reionization for Bayesian analyses that do not include large-scale CMB polarization. The solid green, dashed-dotted purple, and solid blue curves show the Gompertzian $\Lambda$CDM ($\Lambda$G), $\Lambda$CDM without EoR data ($\Lambda$G*), and $w_0w_a$CDM ($w_0w_a$G) constraints, respectively. For comparison, we also show the corresponding $\Lambda$CDM ($\Lambda$T) and $w_0w_a$CDM ($w_0w_a$T) constraints from Ref. \cite{2026PhRvL.136h1002S} as the dashed orange and dashed magenta curves, respectively. The shaded pink region indicates the $1\sigma$ constraint from the ACT DR6 analysis \cite{2025JCAP...11..062L}, which includes low-$\ell$ CMB polarization data.}
    \label{fig:pos_tau}
\end{figure}

Figure \ref{fig:pos_tau} shows the posterior distributions of $\tau_{\rm reio}$ for the different models. Overall, we find a preference for lower values of $\tau_{\rm reio}$ than those reported by \cite{2026PhRvL.136h1002S}, reflecting the use of a physically motivated and observationally supported reionization model. In contrast to the nearly symmetric narrowing of the posterior from $\Lambda$T to $w_0w_a$T, the transition from $\Lambda$G to $w_0w_a$G exhibits a noticeably asymmetric narrowing driven by the inclusion of EoR observations. As discussed in Section~\ref{sec:w0wa}, this behavior ultimately arises from the interplay between constraints provided by CMB lensing and EoR observations.

In the absence of large-scale CMB polarization data, CMB lensing helps break the $A_{\rm s}e^{-2\tau_{\rm reio}}$ degeneracy between the amplitude of the primordial power spectrum $A_{\rm s}$ and $\tau_{\rm reio}$ present in the high-$\ell$ CMB temperature and polarization power spectra. CMB lensing constrains the combination $\sigma_8 \Omega_{\rm m}^{0.25}$, where $\sigma_8$ is the present-day root-mean-square amplitude of the matter density field smoothed over scales of $8\, h^{-1}$ Mpc\footnote{For example, Planck PR3 gives $\sigma_8 \Omega_{\rm m}^{0.25} = 0.589 \pm 0.020$ \cite{2020A&A...641A...8P}.}. Consequently, it also tightly constrains the combination of $A_{\rm s}$ and $\Omega_{\rm m}$.

Interestingly, our model that does not include EoR observations has the same central value as the $w_0w_a$T model. The $\Lambda$G* posterior shape is symmetric given that it does not have the additional constraints at the low-end from, particularly, dark pixel upper limits. The inclusion of EoR measurements breaks the symmetry in the posterior and shifts $\tau_{\rm reio}$ to smaller values than those obtained from $\Lambda$T and $\Lambda$G*. Note that the broad symmetric posteriors of $\Lambda$T and $\Lambda$G* are in good agreement, the usage of a physically motivated model does shifts the posterior to slightly smaller values while also reducing the spread slightly. In terms of cosmological parameters, the shift is due to $\Lambda$T yielding larger $A_{\rm s}$ than the two other $\Lambda$CDM models, which requires larger $\tau_{\rm reio}$ to comply with high-$\ell$ CMB data. 

From our $\Lambda$CDM MCMC analysis, we  obtain the following constraints on $\tau_{\rm reio}$ 
\begin{align}
    \label{eq:tau_LG*}
    &\tau_{\rm reio} = 0.087 \pm 0.011 \ \ \ \ (\mathrm{68\% \ CL, \Lambda G*)}\, , \\
    \label{eq:tau_LG}
    &\tau_{\rm reio} = 0.077 \pm 0.011 \ \ \ \ (\mathrm{68\% \ CL, \Lambda G)}\, . 
\end{align}
For comparison, Ref.~\cite{2026PhRvL.136h1002S} reported $\tau_{\rm reio} = 0.090 \pm 0.012$ (68\% CL) for the $\Lambda$T model. Thus, in $\Lambda$CDM at least, EoR observations still allow relatively large values of the optical depth but with the $0.09$ value outside the $1\sigma$ limit. In the $w_0w_a$CDM scenario, which provides a better fit to the data, the situation changes drastically (see Section \ref{sec:w0wa}).

Refs.~\cite{2025arXiv250821069E,2026arXiv260109644K} also investigated the impact of including EoR neutral hydrogen observations while excluding large-scale CMB polarization data. However, both studies incorporated a broad compilation of reionization constraints, including measurements with varying levels of robustness\footnote{For example, galaxy damping wing constraints may be significantly affected by cosmic variance \cite{2023ApJ...953...29B}.} and model dependence, and employed a pragmatic reconstruction of $x_\HI$ directly from the observations. In contrast, we restrict our analysis to measurements that we consider the most robust, based on their conservative uncertainty estimates, precise redshift determinations, and minimal model dependence. These include quasar damping wing measurements (e.g., \cite{2024MNRAS.530.3208G}) and Ly$\alpha$ (and Ly$\beta$) dark pixel constraints \cite{2015MNRAS.447..499M, 2026MNRAS.545f1862D}. In addition, we adopt the physically motivated Gompertzian reionization model, which was calibrated against reionization simulations \cite{2024arXiv240513680M}, rather than a generic functional form.  

For further comparison, Ref.~\cite{2025arXiv250821069E} reported $\tau_{\rm reio} = 0.0492^{+0.0014}_{-0.0030}$ (68\% CL, $x_\HI (z)$ + BBN + BAO, $\Lambda$CDM), consistent with a rapid and late reionization history while Ref.~\cite{2026arXiv260109644K} found $\tau_{\rm reio} = 0.0552^{+0.0019}_{-0.0026}$ (68\% CL, CMB without lowE + $x_\HI (z)$, $\Lambda$CDM). Our more conservative selection of EoR constraints -- with a physically-motivated reionization model -- yields larger values of $\tau_{\rm reio}$, although they remain below the $\tau_{\rm reio} \simeq 0.08$ required to substantially relax the $\Lambda$CDM$-w_0w_a$CDM tension \cite{2026PhRvL.136h1002S}.

\begin{figure}[H]
    \centering
    \includegraphics[width=0.95\columnwidth]{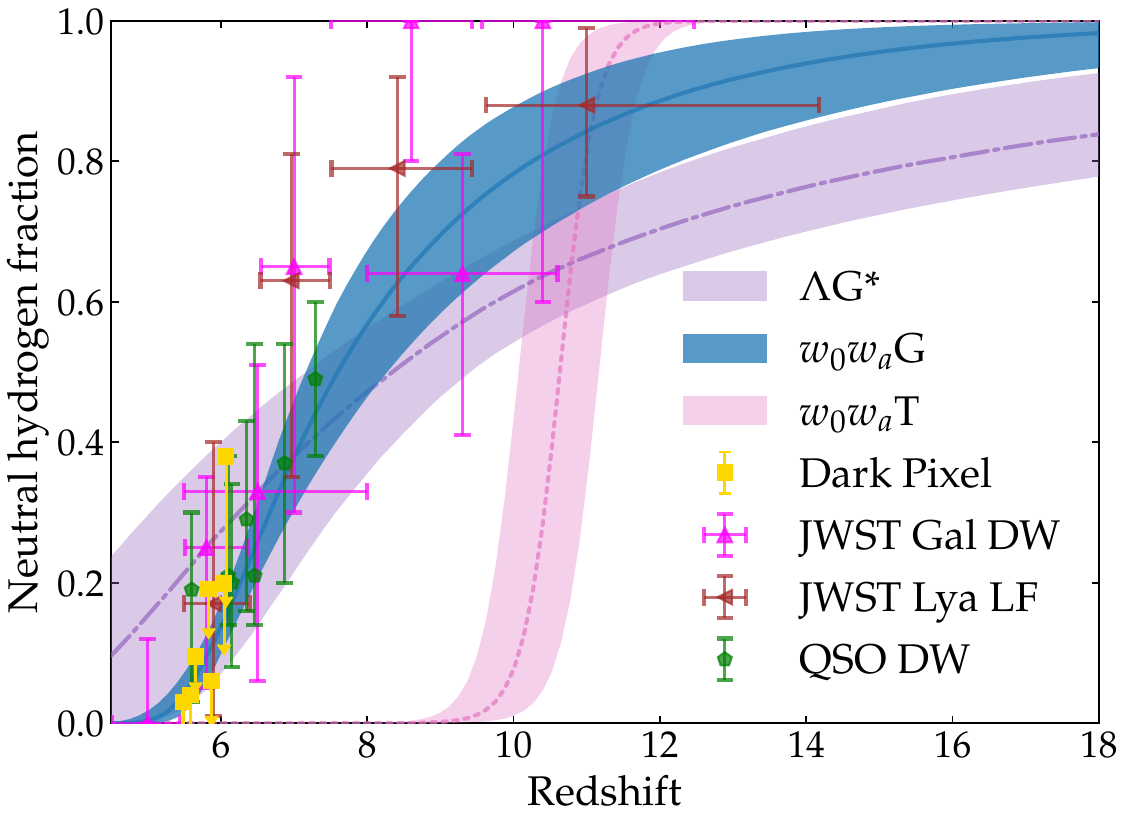}
    \caption{Same as Figure \ref{fig:xHI_LCDM} but for the histories inferred from the $w_0w_a$ scenarios using high-$\ell$ CMB, low-$\ell$ CMB TT, CMB lensing, and BAO as the cosmological dataset. The blue shaded region depicts the reionization history obtained from a Gompertzian reionization model in $w_0w_a$CDM ($w_0w_a$G) with the additional inclusion of  QSO DW and Dark Pixel measurements of the reionization timeline. The magenta shaded region shows the corresponding $1\sigma$ constraint obtained by Ref.~\cite{2026PhRvL.136h1002S} using a hyperbolic tangent ($w_0w_a$T). For comparison, we have included the $\Lambda$G* constraint from Figure \ref{fig:xHI_LCDM} as a benchmark. }
    \label{fig:xHI}
\end{figure}

\section{Preference for dynamical dark energy}
\label{sec:w0wa}
The flexibility of the Gompertzian reionization model allows for a gradual start of reionization followed by a more rapid late-time evolution, in agreement with the behavior predicted by reionization simulations \cite{2011MNRAS.411..955M,2022MNRAS.511.4005K,2023MNRAS.519.6162P,2026arXiv260515310Z}. The integrand of the optical depth favorably weights high-redshift contributions to the timeline of reionization, a positive sign for raising the value of $\tau_{\rm reio}$. However, our analysis also includes EoR observations that constrain the later stages of reionization. As shown in Figure \ref{fig:xHI}, the $1\sigma$ $w_0w_a$G reionization history is already in excellent agreement with high-redshift constraints. Thus, the inclusion of additional, currently less accurate, high-redshift EoR observations is unlikely to increase the inferred value of $\tau_{\rm reio}$ significantly.  

The $w_0w_a$G model yields
\begin{equation}
    \label{eq:tau_w0waG}
    \tau_{\rm reio} = 0.067 \pm 0.011 \ \ \ \ (\mathrm{68\% \ CL, w_0w_a G)}\, ,
\end{equation}
in excellent agreement with the ACT DR6 analysis \cite{2025JCAP...11..062L}, despite excluding low-$\ell$ CMB polarization data. For comparison, although recall that $w_0w_a$T leverages an informative prior centered on high values of the optical depth, Ref.~\cite{2026PhRvL.136h1002S} reported $\tau_{\rm reio} = 0.087 \pm 0.0056$ (68\% CL) for $w_0w_a$T.  

Figure \ref{fig:pos_tau} shows the corresponding posteriors for the optical depth obtained from $w_0w_a$G and $w_0w_a$T. The narrowing of the posterior that occurs from $\Lambda$T to $w_0w_a$T (partially driven by the informative prior on $w_0w_a$T) does not occur, to the same extend, for the Gompertzian reionization models. 

In a $w_0w_a$CDM scenario without EoR data, the additional freedom introduced by $w_0$ and $w_a$ broadens the posterior of $\Omega_{\rm m}$, reconciling early and late probes of $\Omega_{\rm m}$ and ultimately shifting its posterior toward higher values. To satisfy the CMB lensing constraint, $A_{\rm s}$ correspondingly shifts to lower values with its posterior becoming narrower. Through the $A_{\rm s}e^{-2\tau_{\rm reio}}$ degeneracy, this drives $\tau_{\rm reio}$ toward higher values while simultaneously reducing its uncertainty. 

\begin{figure}[H]
    \centering
    \includegraphics[width=0.95\columnwidth]{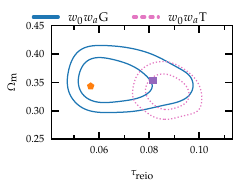}
    \caption{Constraints in the $\Omega_{\rm m} - \tau_{\rm reio}$ plane from the joint analysis of high-$\ell$ CMB, low-$\ell$ CMB TT, CMB lensing, BAO, QSO DW, and dark pixel data using a Gompertzian reionization model ($w_0w_a$G; solid blue contours). The orange pentagon marks the corresponding maximum-likelihood point. For comparison, the dashed magenta contours show the $w_0w_a$CDM results obtained by Ref.~\cite{2026PhRvL.136h1002S} ($w_0w_a$T), with the associated maximum-likelihood point indicated by the purple square.}
    \label{fig:Om_tau}
\end{figure}

The situation changes once EoR observations are included. In this case, $\tau_{\rm reio}$ can no longer increase freely because larger optical depths generally imply earlier reionization histories that could be disfavored by the late stage $x_\HI$ measurements but still allow some leeway for extended timelines. Although CMB lensing stills favors lower values of $A_{\rm s}$, its posterior no longer contracts appreciably, as it must remain consistent with the tighter EoR constraints on $\tau_{\rm reio}$ through their degeneracy.  

Figure \ref{fig:Om_tau} illustrates the relationship between $\tau_{\rm reio}$ and $\Omega_{\rm m}$ for the $w_0w_a$G model. Consistent with the asymmetric posterior shown in Figure \ref{fig:pos_tau}, the $1\sigma$ and $2\sigma$ confidence regions exhibit an extended tail toward larger values of $\tau_{\rm reio}$. As expected, allowing for a dynamical dark energy equation of state substantially weakens the constraint on $\Omega_{\rm m}$, broadening it from $0.2991 \pm 0.0037$ in $\Lambda$G to $0.352 \pm 0.024$ in $w_0w_a$G. A comparable degradation is observed for the hyperbolic tangent models, from $0.2993 \pm 0.0038$ in $\Lambda$T to $0.334 \pm 0.020$ in $w_0w_a$T. 

Figure \ref{fig:Om_tau} also indicates the maximum likelihood points of the $w_0w_a$G and $w_0w_a$T chains. The $w_0w_a$T best-fit solution favors a relatively high value of $\tau_{\rm reio}$, driven in part by the informative prior used in \cite{2026PhRvL.136h1002S}. In contrast, the $w_0w_a$G best-fit solution prefers a lower value of $\tau_{\rm reio}$ that is consistent with analysis including large-scale CMB polarization.

\begin{figure}[H]
    \centering
    \includegraphics[width=0.95\columnwidth]{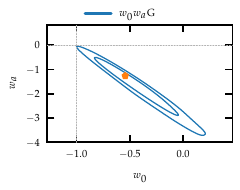}
    \caption{Constraints in the $w_0 - w_a$ plane from the joint analysis of epoch of reionization observations, low-$\ell$ CMB TT, high-$\ell$ CMB, CMB lensing, and BAO using a Gompertzian reionization model. The contours denote the 68\% and 95\% confidence levels, while the orange pentagon marks the maximum-likelihood point of the MCMC posterior.}
    \label{fig:w0wa}
\end{figure}

Figure \ref{fig:w0wa} shows that a joint analysis combining low-$\ell$ CMB TT, high-$\ell$ CMB TTTEEE, CMB lensing, BAO, QSO DW, and dark pixel $x_\HI$ constraints, together with a physically motivated reionization model, still yields a preference for $w_0w_a$CDM with $\Lambda$CDM excluded at $\gtrapprox 2\sigma$. Moreover, the maximum likelihood point is located decisively away from the $\Lambda$CDM point. 

Specifically, we obtain the following 68\% CL constraints on the dynamical dark energy parameters
\begin{align}
    w_0 &= -0.42 \pm 0.25 \, ,\\
    w_a &= -1.74 \pm 0.77 \, ,
\end{align}
while the corresponding 95\% CL upper bound on the sum of the neutrino masses is $\sum m_\nu < 0.219$ eV. The posterior peaks at approximately $0.1$ eV, making it compatible with both the normal and inverted neutrino mass hierarchies and in excellent agreement with neutrino oscillation experiments \cite{2024JHEP...12..216E}. We note that the inclusion of large-angular polarization in the Gompertzian framework further shifts the optical depth to smaller values ($\tau_{\rm reio} \approx 0.058$) and further strengthens the $w_0w_a$CDM$-\Lambda$CDM tension to $>2\sigma$ level \cite{2026arXiv260413423M}.

We find $\Delta \chi^2 \equiv \chi^2_{w_0w_a\mathrm{G}} - \chi^2_{\Lambda\mathrm{G}} = -4.6$, indicating a preference for the dynamical dark energy scenario. However, the improvement in the fit is modest given the additional complexity of the model due to the increased number of parameters introduced in the $w_0w_a$G model.

\section{Conclusions}
\label{sec:conclusion}
We present a joint analysis of CMB, BAO, and reionization observations excluding large-scale CMB polarization data. Our analysis adopts a Gompertzian reionization model \cite{2024arXiv240513680M}, a physically motivated framework calibrated against reionization simulations and designed to jointly fit CMB and reionization observations \cite{2026arXiv260413423M}. 

We find that including quasar damping wing and dark pixel constraints on $x_\HI$ yields larger values of $\tau_{\rm reio}$ than those reported in \cite{2025arXiv250821069E,2026arXiv260109644K}, which reconstructed $x_\HI(z)$ directly from a heterogeneous compilation of EoR observations. Nevertheless, in our analysis the optical depth decreases when extending $\Lambda$CDM to $w_0w_a$CDM, with $\Delta \tau_{\rm reio} = -0.010$. Interestingly, this trend is also present in other works. For instance, Ref.~\cite{2026PhRvL.136h1002S} results yield $\Delta \tau_{\rm reio} = -0.003$, while Ref.~\cite{2025arXiv250821069E} likewise found a decrease in $\tau_{\rm reio}$ when moving from $\Lambda$CDM to $w_0w_a$CDM, although the shift was small ($\Delta \tau_{\rm reio} = - 0.0006$) for their combination of BAO, BBN, and a broad compilation of $x_\HI$ measurements. 

Our $\Lambda$CDM analysis excludes high values of the optical depth to reionization ($\tau_{\rm reio} \approx 0.09$) at more than $1\sigma$. Moreover, we found that the dynamical dark energy model provides a better fit to the data. 

Notably, our $w_0w_a$CDM analysis yields a value of $\tau_{\rm reio}$ that is consistent with analyses including large-scale CMB polarization data (see Figure \ref{fig:pos_tau}), while simultaneously providing an excellent match to multiple independent measurements of the reionization history that were not included in the likelihood computation (see Figure \ref{fig:xHI}). Despite excluding large-scale CMB polarization data, incorporating quasar damping wing and dark pixel $x_\HI$ constraints together with a physically motivated reionization model still leaves a $\gtrapprox 2\sigma$ preference for $w_0w_a$CDM over $\Lambda$CDM (see Figure \ref{fig:w0wa}). Finally, although the increase in $\tau_{\rm reio}$ is insufficient to substantially alleviate the dark energy tension, it is large enough to resolve the apparent preference for a negative sum of neutrino masses.

We conclude that, even if an undiscovered systematic were present in large-scale CMB polarization spectra, it would not by itself restore consistency with $\Lambda$CDM. Reconciling $\Lambda$CDM with the data would also require multiple independent EoR observations to be affected by previously unidentified systematics in a mutually consistent manner. Nevertheless, relaxing some of the physical assumptions underlying $\Lambda$CDM could help alleviate the remaining tension within our framework. For instance, departing from the cold dark matter paradigm introduces additional freedom to modify the energy injected into the intergalactic medium during cosmic dawn and the early stages of reionization. In particular, dark matter annihilation, dark matter decay, and primordial black holes can inject energy into the IGM, partially ionizing neutral hydrogen before the onset of astrophysical reionization \cite{2024PhRvD.110l3529X,2025PhRvD.112j3048C,2025JCAP...04..081H,2026JCAP...02..060K,2026PhRvD.113f3031K}, hence modifying the early evolution of $x_\HI(z)$. Such scenarios are naturally compatible with the Gompertzian framework and will be explored in future work.

\Acknowledgements{
The authors thank Noah Sailer for comments. \\
PMC acknowledges the support from the Basic and Frontier Research Project of PCL (grant No.~2025QYB012), and the Major Key Project of PCL. YM acknowledges the support from the National SKA Program of China (grant No. 2020SKA0110401). The authors acknowledge the Tsinghua Astrophysics High-Performance Computing platform at Tsinghua University and PCL's Cloud Brain for providing computational and data storage resources that have contributed to the research results reported within this paper.}

\InterestConflict{
    The authors declare that they have no conflict of interest.
}

\bibliography{main}

@ARTICLE{2026arXiv260413423M,
       author = {{Montero-Camacho}, Paulo and {Li}, Yin and {Barquero-Hern{\'a}ndez}, Marco and {Renard}, Pablo and {Wang}, Yang and {Li}, Xiao-Dong},
        title = "{Into the Gompverse: A robust Gompertzian reionization model for CMB analyses}",
      journal = {arXiv e-prints},
         year = 2026,
        month = apr,
          eid = {arXiv:2604.13423},
        pages = {arXiv:2604.13423},
          doi = {10.48550/arXiv.2604.13423},
archivePrefix = {arXiv},
       eprint = {2604.13423},
 primaryClass = {astro-ph.CO},
       adsurl = {https://ui.adsabs.harvard.edu/abs/2026arXiv260413423M}
}

@ARTICLE{2024arXiv240513680M,
       author = {{Montero-Camacho}, Paulo and {Li}, Yin and {Cranmer}, Miles},
        title = "{Five parameters are all you need (in $\Lambda$CDM)}",
      journal = {arXiv e-prints},
         year = 2024,
        month = may,
          eid = {arXiv:2405.13680},
        pages = {arXiv:2405.13680},
          doi = {10.48550/arXiv.2405.13680},
archivePrefix = {arXiv},
       eprint = {2405.13680},
 primaryClass = {astro-ph.CO},
       adsurl = {https://ui.adsabs.harvard.edu/abs/2024arXiv240513680M}
}

@ARTICLE{2025PhRvD.112d3506F,
       author = {{Facchinetti}, Ga{\'e}tan},
        title = "{Neural network emulation of reionization to constrain new physics with early- and late-time probes}",
      journal = {\prd},
         year = 2025,
        month = aug,
       volume = {112},
       number = {4},
          eid = {043506},
        pages = {043506},
          doi = {10.1103/j92y-wbtn},
archivePrefix = {arXiv},
       eprint = {2503.11261},
 primaryClass = {astro-ph.CO},
       adsurl = {https://ui.adsabs.harvard.edu/abs/2025PhRvD.112d3506F}
}

@ARTICLE{2016A&A...596A.108P,
       author = {{Planck Collaboration} and {Adam}, R. and others},
        title = "{Planck intermediate results. XLVII. Planck constraints on reionization history}",
      journal = {\aap},
         year = 2016,
        month = dec,
       volume = {596},
          eid = {A108},
        pages = {A108},
          doi = {10.1051/0004-6361/201628897},
archivePrefix = {arXiv},
       eprint = {1605.03507},
 primaryClass = {astro-ph.CO},
       adsurl = {https://ui.adsabs.harvard.edu/abs/2016A&A...596A.108P}
}

@ARTICLE{2015A&A...580L...4D,
       author = {{Douspis}, M. and {Aghanim}, N. and {Ili{\'c}}, S. and {Langer}, M.},
        title = "{A new parameterization of the reionisation history}",
      journal = {\aap},
         year = 2015,
        month = aug,
       volume = {580},
          eid = {L4},
        pages = {L4},
          doi = {10.1051/0004-6361/201526543},
archivePrefix = {arXiv},
       eprint = {1509.02785},
 primaryClass = {astro-ph.CO},
       adsurl = {https://ui.adsabs.harvard.edu/abs/2015A&A...580L...4D}
}

@ARTICLE{2018ApJ...858L..11T,
       author = {{Trac}, Hy},
        title = "{Parametrizing the Reionization History with the Redshift Midpoint, Duration, and Asymmetry}",
      journal = {\apjl},
         year = 2018,
        month = may,
       volume = {858},
       number = {2},
          eid = {L11},
        pages = {L11},
          doi = {10.3847/2041-8213/aabff0},
archivePrefix = {arXiv},
       eprint = {1804.00672},
 primaryClass = {astro-ph.CO},
       adsurl = {https://ui.adsabs.harvard.edu/abs/2018ApJ...858L..11T}
}

@ARTICLE{2026PhRvL.136h1002S,
       author = {{Sailer}, Noah and {Farren}, Gerrit S. and {Ferraro}, Simone and {White}, Martin},
        title = "{Addressing Tensions in {\ensuremath{\Lambda}}CDM Cosmology by an Increase in the Optical Depth to Reionization}",
      journal = {\prl},
         year = 2026,
        month = feb,
       volume = {136},
       number = {8},
          eid = {081002},
        pages = {081002},
          doi = {10.1103/6r54-8lv4},
archivePrefix = {arXiv},
       eprint = {2504.16932},
 primaryClass = {astro-ph.CO},
       adsurl = {https://ui.adsabs.harvard.edu/abs/2026PhRvL.136h1002S}
}

@ARTICLE{2025PhRvD.112d3541J,
       author = {{Jhaveri}, Tanisha and {Karwal}, Tanvi and {Hu}, Wayne},
        title = "{Turning a negative neutrino mass into a positive optical depth}",
      journal = {\prd},
         year = 2025,
        month = aug,
       volume = {112},
       number = {4},
          eid = {043541},
        pages = {043541},
          doi = {10.1103/6vd2-rbfn},
archivePrefix = {arXiv},
       eprint = {2504.21813},
 primaryClass = {astro-ph.CO},
       adsurl = {https://ui.adsabs.harvard.edu/abs/2025PhRvD.112d3541J}
}

@ARTICLE{2025ApJ...983L..27T,
       author = {{Tang}, Xianzhe TZ and {Brout}, Dillon and {Karwal}, Tanvi and {Chang}, Chihway and {Miranda}, Vivian and {Vincenzi}, Maria},
        title = "{Uniting the Observed Dynamical Dark Energy Preference with the Discrepancies in {\ensuremath{\Omega}}$_{m}$ and H$_{0}$ across Cosmological Probes}",
      journal = {\apjl},
         year = 2025,
        month = apr,
       volume = {983},
       number = {1},
          eid = {L27},
        pages = {L27},
          doi = {10.3847/2041-8213/adc4da},
archivePrefix = {arXiv},
       eprint = {2412.04430},
 primaryClass = {astro-ph.CO},
       adsurl = {https://ui.adsabs.harvard.edu/abs/2025ApJ...983L..27T}
}

@ARTICLE{2025PhRvD.112h3515A,
       author = {{DESI Collaboration} and {Abdul Karim}, M. and others},
        title = "{DESI DR2 results. II. Measurements of baryon acoustic oscillations and cosmological constraints}",
      journal = {\prd},
         year = 2025,
        month = oct,
       volume = {112},
       number = {8},
          eid = {083515},
        pages = {083515},
          doi = {10.1103/tr6y-kpc6},
archivePrefix = {arXiv},
       eprint = {2503.14738},
 primaryClass = {astro-ph.CO},
       adsurl = {https://ui.adsabs.harvard.edu/abs/2025PhRvD.112h3515A}
}

@ARTICLE{2020A&A...641A...5P,
       author = {{Planck Collaboration} and {Aghanim}, N. and others},
        title = "{Planck 2018 results. V. CMB power spectra and likelihoods}",
      journal = {\aap},
         year = 2020,
        month = sep,
       volume = {641},
          eid = {A5},
        pages = {A5},
          doi = {10.1051/0004-6361/201936386},
archivePrefix = {arXiv},
       eprint = {1907.12875},
 primaryClass = {astro-ph.CO},
       adsurl = {https://ui.adsabs.harvard.edu/abs/2020A&A...641A...5P}
}

@ARTICLE{2025JCAP...11..062L,
       author = {{The Atacama Cosmology Telescope Collaboration} and {Louis}, Thibaut and others},
        title = "{The Atacama Cosmology Telescope: DR6 power spectra, likelihoods and {\ensuremath{\Lambda}}CDM parameters}",
      journal = {\jcap},
         year = 2025,
        month = nov,
       volume = {2025},
       number = {11},
          eid = {062},
        pages = {062},
          doi = {10.1088/1475-7516/2025/11/062},
archivePrefix = {arXiv},
       eprint = {2503.14452},
 primaryClass = {astro-ph.CO},
       adsurl = {https://ui.adsabs.harvard.edu/abs/2025JCAP...11..062L}
}

@ARTICLE{2025PhRvD.112h3513E,
       author = {{Elbers}, W. and others},
        title = "{Constraints on neutrino physics from DESI DR2 BAO and DR1 full shape}",
      journal = {\prd},
         year = 2025,
        month = oct,
       volume = {112},
       number = {8},
          eid = {083513},
        pages = {083513},
          doi = {10.1103/w9pk-xsk7},
archivePrefix = {arXiv},
       eprint = {2503.14744},
 primaryClass = {astro-ph.CO},
       adsurl = {https://ui.adsabs.harvard.edu/abs/2025PhRvD.112h3513E}
}

@ARTICLE{2025PhRvD.111h3507G,
       author = {{Green}, Daniel and {Meyers}, Joel},
        title = "{Cosmological preference for a negative neutrino mass}",
      journal = {\prd},
         year = 2025,
        month = apr,
       volume = {111},
       number = {8},
          eid = {083507},
        pages = {083507},
          doi = {10.1103/PhysRevD.111.083507},
archivePrefix = {arXiv},
       eprint = {2407.07878},
 primaryClass = {astro-ph.CO},
       adsurl = {https://ui.adsabs.harvard.edu/abs/2025PhRvD.111h3507G}
}

@ARTICLE{2020A&A...641A...1P,
       author = {{Planck Collaboration} and {Aghanim}, N. and others},
        title = "{Planck 2018 results. I. Overview and the cosmological legacy of Planck}",
      journal = {\aap},
         year = 2020,
        month = sep,
       volume = {641},
          eid = {A1},
        pages = {A1},
          doi = {10.1051/0004-6361/201833880},
archivePrefix = {arXiv},
       eprint = {1807.06205},
 primaryClass = {astro-ph.CO},
       adsurl = {https://ui.adsabs.harvard.edu/abs/2020A&A...641A...1P}
}

@ARTICLE{2019A&A...629A..38D,
       author = {{Delouis}, J.-M. and {Pagano}, L. and {Mottet}, S. and {Puget}, J.-L. and {Vibert}, L.},
        title = "{SRoll2: an improved mapmaking approach to reduce large-scale systematic effects in the Planck High Frequency Instrument legacy maps}",
      journal = {\aap},
         year = 2019,
        month = sep,
       volume = {629},
          eid = {A38},
        pages = {A38},
          doi = {10.1051/0004-6361/201834882},
archivePrefix = {arXiv},
       eprint = {1901.11386},
 primaryClass = {astro-ph.CO},
       adsurl = {https://ui.adsabs.harvard.edu/abs/2019A&A...629A..38D}
}

@ARTICLE{2021A&A...651A..65L,
       author = {{Lopez-Radcenco}, M. and {Delouis}, J.-M. and {Vibert}, L.},
        title = "{SRoll3: A neural network approach to reduce large-scale systematic effects in the Planck High-Frequency Instrument maps}",
      journal = {\aap},
         year = 2021,
        month = jul,
       volume = {651},
          eid = {A65},
        pages = {A65},
          doi = {10.1051/0004-6361/202040152},
archivePrefix = {arXiv},
       eprint = {2012.09702},
 primaryClass = {astro-ph.IM},
       adsurl = {https://ui.adsabs.harvard.edu/abs/2021A&A...651A..65L}
}

@ARTICLE{2020A&A...635A..99P,
       author = {{Pagano}, L. and {Delouis}, J.-M. and {Mottet}, S. and {Puget}, J.-L. and {Vibert}, L.},
        title = "{Reionization optical depth determination from Planck HFI data with ten percent accuracy}",
      journal = {\aap},
         year = 2020,
        month = mar,
       volume = {635},
          eid = {A99},
        pages = {A99},
          doi = {10.1051/0004-6361/201936630},
archivePrefix = {arXiv},
       eprint = {1908.09856},
 primaryClass = {astro-ph.CO},
       adsurl = {https://ui.adsabs.harvard.edu/abs/2020A&A...635A..99P}
}

@ARTICLE{2026arXiv260218761F,
       author = {{Forero-S{\'a}nchez}, D. and others},
        title = "{Cosmological constraints from a joint DESI DR1 Full-Shape and DR2 BAO}",
      journal = {arXiv e-prints},
         year = 2026,
        month = feb,
          eid = {arXiv:2602.18761},
        pages = {arXiv:2602.18761},
          doi = {10.48550/arXiv.2602.18761},
archivePrefix = {arXiv},
       eprint = {2602.18761},
 primaryClass = {astro-ph.CO},
       adsurl = {https://ui.adsabs.harvard.edu/abs/2026arXiv260218761F}
}

@ARTICLE{2026arXiv260623936F,
       author = {{Forero-S{\'a}nchez}, D. and others},
        title = "{Cosmological constraints from the DESI DR1 Bispectrum Full-Shape and DR2 BAO}",
      journal = {arXiv e-prints},
         year = 2026,
        month = jun,
          eid = {arXiv:2606.23936},
        pages = {arXiv:2606.23936},
          doi = {10.48550/arXiv.2606.23936},
archivePrefix = {arXiv},
       eprint = {2606.23936},
 primaryClass = {astro-ph.CO},
       adsurl = {https://ui.adsabs.harvard.edu/abs/2026arXiv260623936F}
}

@ARTICLE{2022MNRAS.511.4005K,
       author = {{Kannan}, R. and {Garaldi}, E. and {Smith}, A. and {Pakmor}, R. and {Springel}, V. and {Vogelsberger}, M. and {Hernquist}, L.},
        title = "{Introducing the THESAN project: radiation-magnetohydrodynamic simulations of the epoch of reionization}",
      journal = {\mnras},
         year = 2022,
        month = apr,
       volume = {511},
       number = {3},
        pages = {4005-4030},
          doi = {10.1093/mnras/stab3710},
archivePrefix = {arXiv},
       eprint = {2110.00584},
 primaryClass = {astro-ph.GA},
       adsurl = {https://ui.adsabs.harvard.edu/abs/2022MNRAS.511.4005K}
}

@ARTICLE{2011MNRAS.411..955M,
       author = {{Mesinger}, Andrei and {Furlanetto}, Steven and {Cen}, Renyue},
        title = "{21CMFAST: a fast, seminumerical simulation of the high-redshift 21-cm signal}",
      journal = {\mnras},
         year = 2011,
        month = feb,
       volume = {411},
       number = {2},
        pages = {955-972},
          doi = {10.1111/j.1365-2966.2010.17731.x},
archivePrefix = {arXiv},
       eprint = {1003.3878},
 primaryClass = {astro-ph.CO},
       adsurl = {https://ui.adsabs.harvard.edu/abs/2011MNRAS.411..955M}
}

@ARTICLE{2023MNRAS.519.6162P,
       author = {{Puchwein}, Ewald and {Bolton}, James S. and {Keating}, Laura C. and {Molaro}, Margherita and {Gaikwad}, Prakash and {Kulkarni}, Girish and {Haehnelt}, Martin G. and {Ir{\v{s}}i{\v{c}}}, Vid and {{\v{S}}oltinsk{\'y}}, Tom{\'a}{\v{s}} and {Viel}, Matteo and {Aubert}, Dominique and {Becker}, George D. and {Meiksin}, Avery},
        title = "{The Sherwood-Relics simulations: overview and impact of patchy reionization and pressure smoothing on the intergalactic medium}",
      journal = {\mnras},
         year = 2023,
        month = mar,
       volume = {519},
       number = {4},
        pages = {6162-6183},
          doi = {10.1093/mnras/stac3761},
archivePrefix = {arXiv},
       eprint = {2207.13098},
 primaryClass = {astro-ph.CO},
       adsurl = {https://ui.adsabs.harvard.edu/abs/2023MNRAS.519.6162P}
}

@ARTICLE{2026arXiv260515310Z,
       author = {{Zier}, Oliver and {Smith}, Aaron and {Shen}, Xuejian and {Liu}, Rongrong and {Kannan}, Rahul and {Koehler}, Sonja M. and {Springel}, Volker and {Pakmor}, R{\"u}diger and {Vogelsberger}, Mark and {Bulichi}, Teodora-Elena and {Hernquist}, Lars},
        title = "{Introducing the Lumina project: large-volume radiation-hydrodynamic simulations of the epochs of hydrogen and helium reionization}",
      journal = {arXiv e-prints},
         year = 2026,
        month = may,
          eid = {arXiv:2605.15310},
        pages = {arXiv:2605.15310},
          doi = {10.48550/arXiv.2605.15310},
archivePrefix = {arXiv},
       eprint = {2605.15310},
 primaryClass = {astro-ph.CO},
       adsurl = {https://ui.adsabs.harvard.edu/abs/2026arXiv260515310Z}
}

@ARTICLE{2026arXiv260620795J,
       author = {{Jhaveri}, Tanisha and {Hu}, Wayne and {Miranda}, Vivian},
        title = "{Raising the reionization optical depth with inflationary CMB features}",
      journal = {arXiv e-prints},
         year = 2026,
        month = jun,
          eid = {arXiv:2606.20795},
        pages = {arXiv:2606.20795},
          doi = {10.48550/arXiv.2606.20795},
archivePrefix = {arXiv},
       eprint = {2606.20795},
 primaryClass = {astro-ph.CO},
       adsurl = {https://ui.adsabs.harvard.edu/abs/2026arXiv260620795J}
}

@ARTICLE{2026arXiv260619459A,
       author = {{Aggarwal}, Yash and {Cain}, Christopher and {Lopez}, Garett and {Trac}, Hy and {D'Aloisio}, Anson and {Tanedo}, Philip and {Tan}, Jonathan C.},
        title = "{Fireworks at Cosmic Dawn: relieving BAO-CMB tensions with the Pop III.1 Flash}",
      journal = {arXiv e-prints},
         year = 2026,
        month = jun,
          eid = {arXiv:2606.19459},
        pages = {arXiv:2606.19459},
          doi = {10.48550/arXiv.2606.19459},
archivePrefix = {arXiv},
       eprint = {2606.19459},
 primaryClass = {astro-ph.CO},
       adsurl = {https://ui.adsabs.harvard.edu/abs/2026arXiv260619459A}
}

@ARTICLE{2022AJ....164..207D,
       author = {{DESI Collaboration} and {Abareshi}, B. and others},
        title = "{Overview of the Instrumentation for the Dark Energy Spectroscopic Instrument}",
      journal = {\aj},
         year = 2022,
        month = nov,
       volume = {164},
       number = {5},
          eid = {207},
        pages = {207},
          doi = {10.3847/1538-3881/ac882b},
archivePrefix = {arXiv},
       eprint = {2205.10939},
 primaryClass = {astro-ph.IM},
       adsurl = {https://ui.adsabs.harvard.edu/abs/2022AJ....164..207D}
}

@ARTICLE{2019arXiv191000483E,
       author = {{Efstathiou}, George and {Gratton}, Steven},
        title = "{A Detailed Description of the CamSpec Likelihood Pipeline and a Reanalysis of the Planck High Frequency Maps}",
      journal = {arXiv e-prints},
         year = 2019,
        month = oct,
          eid = {arXiv:1910.00483},
        pages = {arXiv:1910.00483},
          doi = {10.48550/arXiv.1910.00483},
archivePrefix = {arXiv},
       eprint = {1910.00483},
 primaryClass = {astro-ph.CO},
       adsurl = {https://ui.adsabs.harvard.edu/abs/2019arXiv191000483E}
}

@ARTICLE{2024ApJ...962..112Q,
       author = {{Qu}, Frank J. and others},
        title = "{The Atacama Cosmology Telescope: A Measurement of the DR6 CMB Lensing Power Spectrum and Its Implications for Structure Growth}",
      journal = {\apj},
         year = 2024,
        month = feb,
       volume = {962},
       number = {2},
          eid = {112},
        pages = {112},
          doi = {10.3847/1538-4357/acfe06},
archivePrefix = {arXiv},
       eprint = {2304.05202},
 primaryClass = {astro-ph.CO},
       adsurl = {https://ui.adsabs.harvard.edu/abs/2024ApJ...962..112Q}
}

@ARTICLE{2022JCAP...09..039C,
       author = {{Carron}, Julien and {Mirmelstein}, Mark and {Lewis}, Antony},
        title = "{CMB lensing from Planck PR4 maps}",
      journal = {\jcap},
         year = 2022,
        month = sep,
       volume = {2022},
       number = {9},
          eid = {039},
        pages = {039},
          doi = {10.1088/1475-7516/2022/09/039},
archivePrefix = {arXiv},
       eprint = {2206.07773},
 primaryClass = {astro-ph.CO},
       adsurl = {https://ui.adsabs.harvard.edu/abs/2022JCAP...09..039C}
}

@ARTICLE{2026MNRAS.545f1862D,
       author = {{Davies}, Frederick B. and {Bosman}, Sarah E.~I. and {D'Odorico}, Valentina and {Campo}, Sofia and {Mesinger}, Andrei and {Qin}, Yuxiang and {Becker}, George D. and {Ba{\~n}ados}, Eduardo and {Chen}, Huanqing and {Cristiani}, Stefano and {Fan}, Xiaohui and {Gallerani}, Simona and {Haehnelt}, Martin G. and {Keating}, Laura C. and {Lai}, Samuel and {Ryan-Weber}, Emma and {Wang}, Feige and {Yang}, Jinyi and {Zhu}, Yongda},
        title = "{Updated dark pixel fraction constraints on reionization's end from the Lyman-series forests of XQR{\ensuremath{-}}30}",
      journal = {\mnras},
         year = 2026,
        month = jan,
       volume = {545},
       number = {2},
          eid = {staf1862},
        pages = {staf1862},
          doi = {10.1093/mnras/staf1862},
archivePrefix = {arXiv},
       eprint = {2510.25829},
 primaryClass = {astro-ph.CO},
       adsurl = {https://ui.adsabs.harvard.edu/abs/2026MNRAS.545f1862D}
}

@ARTICLE{2022MNRAS.512.5390G,
       author = {{Greig}, Bradley and {Mesinger}, Andrei and {Davies}, Frederick B. and {Wang}, Feige and {Yang}, Jinyi and {Hennawi}, Joseph F.},
        title = "{IGM damping wing constraints on reionization from covariance reconstruction of two z {\ensuremath{\gtrsim}} 7 QSOs}",
      journal = {\mnras},
         year = 2022,
        month = jun,
       volume = {512},
       number = {4},
        pages = {5390-5403},
          doi = {10.1093/mnras/stac825},
archivePrefix = {arXiv},
       eprint = {2112.04091},
 primaryClass = {astro-ph.CO},
       adsurl = {https://ui.adsabs.harvard.edu/abs/2022MNRAS.512.5390G}
}

@ARTICLE{2024MNRAS.530.3208G,
       author = {{Greig}, B. and {Mesinger}, A. and {Ba{\~n}ados}, E. and {Becker}, G.~D. and {Bosman}, S.~E.~I. and {Chen}, H. and {Davies}, F.~B. and {D'Odorico}, V. and {Eilers}, A. -C. and {Gallerani}, S. and {Haehnelt}, M.~G. and {Keating}, L. and {Lai}, S. and {Qin}, Y. and {Ryan-Weber}, E. and {Satyavolu}, S. and {Wang}, F. and {Yang}, J. and {Zhu}, Y.},
        title = "{IGM damping wing constraints on the tail end of reionization from the enlarged XQR-30 sample}",
      journal = {\mnras},
         year = 2024,
        month = may,
       volume = {530},
       number = {3},
        pages = {3208-3227},
          doi = {10.1093/mnras/stae1080},
archivePrefix = {arXiv},
       eprint = {2404.12585},
 primaryClass = {astro-ph.CO},
       adsurl = {https://ui.adsabs.harvard.edu/abs/2024MNRAS.530.3208G}
}

@ARTICLE{2024A&A...688L..26S,
       author = {{Spina}, Benedetta and {Bosman}, Sarah E.~I. and {Davies}, Frederick B. and {Gaikwad}, Prakash and {Zhu}, Yongda},
        title = "{Damping wings in the Lyman-{\ensuremath{\alpha}} forest: A model-independent measurement of the neutral fraction at 5.4 < z < 6.1}",
      journal = {\aap},
         year = 2024,
        month = aug,
       volume = {688},
          eid = {L26},
        pages = {L26},
          doi = {10.1051/0004-6361/202450798},
archivePrefix = {arXiv},
       eprint = {2405.12273},
 primaryClass = {astro-ph.CO},
       adsurl = {https://ui.adsabs.harvard.edu/abs/2024A&A...688L..26S}
}

@ARTICLE{2024ApJ...969..162D,
       author = {{{\v{D}}urov{\v{c}}{\'\i}kov{\'a}}, Dominika and {Eilers}, Anna-Christina and {Chen}, Huanqing and {Satyavolu}, Sindhu and {Kulkarni}, Girish and {Simcoe}, Robert A. and {Keating}, Laura C. and {Haehnelt}, Martin G. and {Ba{\~n}ados}, Eduardo},
        title = "{Chronicling the Reionization History at 6 {\ensuremath{\lesssim}} z {\ensuremath{\lesssim}} 7 with Emergent Quasar Damping Wings}",
      journal = {\apj},
         year = 2024,
        month = jul,
       volume = {969},
       number = {2},
          eid = {162},
        pages = {162},
          doi = {10.3847/1538-4357/ad4888},
archivePrefix = {arXiv},
       eprint = {2401.10328},
 primaryClass = {astro-ph.CO},
       adsurl = {https://ui.adsabs.harvard.edu/abs/2024ApJ...969..162D}
}

@ARTICLE{2025arXiv250821069E,
       author = {{Elbers}, Willem},
        title = "{Rapid late-time reionization: constraints and cosmological implications}",
      journal = {arXiv e-prints},
         year = 2025,
        month = aug,
          eid = {arXiv:2508.21069},
        pages = {arXiv:2508.21069},
          doi = {10.48550/arXiv.2508.21069},
archivePrefix = {arXiv},
       eprint = {2508.21069},
 primaryClass = {astro-ph.CO},
       adsurl = {https://ui.adsabs.harvard.edu/abs/2025arXiv250821069E}
}

@ARTICLE{2026arXiv260109644K,
       author = {{Kageura}, Yuta and {Ouchi}, Masami and {Naokawa}, Fumihiro and {Umeda}, Hiroya and {Matsumoto}, Akinori and {Harikane}, Yuichi and {Nakane}, Minami and {Thai}, Tran Thi},
        title = "{A New Constraint on the Optical Depth from the Reionization History Independent of CMB Large-Scale E-Mode Polarization}",
      journal = {arXiv e-prints},
         year = 2026,
        month = jan,
          eid = {arXiv:2601.09644},
        pages = {arXiv:2601.09644},
          doi = {10.48550/arXiv.2601.09644},
archivePrefix = {arXiv},
       eprint = {2601.09644},
 primaryClass = {astro-ph.CO},
       adsurl = {https://ui.adsabs.harvard.edu/abs/2026arXiv260109644K}
}

@ARTICLE{2023ApJ...953...29B,
       author = {{Bruton}, Sean and {Scarlata}, Claudia and {Haardt}, Francesco and {Hayes}, Matthew J. and {Mason}, Charlotte and {Morales}, Alexa M. and {Mesinger}, Andrei},
        title = "{The Impact of Cosmic Variance on Inferences of Global Neutral Fraction Derived from Ly{\ensuremath{\alpha}} Luminosity Functions during Reionization}",
      journal = {\apj},
         year = 2023,
        month = aug,
       volume = {953},
       number = {1},
          eid = {29},
        pages = {29},
          doi = {10.3847/1538-4357/acd179},
archivePrefix = {arXiv},
       eprint = {2305.04949},
 primaryClass = {astro-ph.CO},
       adsurl = {https://ui.adsabs.harvard.edu/abs/2023ApJ...953...29B}
}

@ARTICLE{2021JCAP...05..057T,
       author = {{Torrado}, Jes{\'u}s and {Lewis}, Antony},
        title = "{Cobaya: code for Bayesian analysis of hierarchical physical models}",
      journal = {\jcap},
         year = 2021,
        month = may,
       volume = {2021},
       number = {5},
          eid = {057},
        pages = {057},
          doi = {10.1088/1475-7516/2021/05/057},
archivePrefix = {arXiv},
       eprint = {2005.05290},
 primaryClass = {astro-ph.IM},
       adsurl = {https://ui.adsabs.harvard.edu/abs/2021JCAP...05..057T}
}

@ARTICLE{2011JCAP...07..034B,
       author = {{Blas}, Diego and {Lesgourgues}, Julien and {Tram}, Thomas},
        title = "{The Cosmic Linear Anisotropy Solving System (CLASS). Part II: Approximation schemes}",
      journal = {\jcap},
         year = 2011,
        month = jul,
       volume = {2011},
       number = {7},
          eid = {034},
        pages = {034},
          doi = {10.1088/1475-7516/2011/07/034},
archivePrefix = {arXiv},
       eprint = {1104.2933},
 primaryClass = {astro-ph.CO},
       adsurl = {https://ui.adsabs.harvard.edu/abs/2011JCAP...07..034B}
}

@ARTICLE{2020A&A...641A...8P,
       author = {{Planck Collaboration} and {Aghanim}, N. and {Akrami}, Y. and {Ashdown}, M. and {Aumont}, J. and {Baccigalupi}, C. and {Ballardini}, M. and {Banday}, A.~J. and {Barreiro}, R.~B. and {Bartolo}, N. and {Basak}, S. and {Benabed}, K. and {Bernard}, J.-P. and {Bersanelli}, M. and {Bielewicz}, P. and {Bock}, J.~J. and {Bond}, J.~R. and {Borrill}, J. and {Bouchet}, F.~R. and {Boulanger}, F. and {Bucher}, M. and {Burigana}, C. and {Calabrese}, E. and {Cardoso}, J.-F. and {Carron}, J. and {Challinor}, A. and {Chiang}, H.~C. and {Colombo}, L.~P.~L. and {Combet}, C. and {Crill}, B.~P. and {Cuttaia}, F. and {de Bernardis}, P. and {de Zotti}, G. and {Delabrouille}, J. and {Di Valentino}, E. and {Diego}, J.~M. and {Dor{\'e}}, O. and {Douspis}, M. and {Ducout}, A. and {Dupac}, X. and {Efstathiou}, G. and {Elsner}, F. and {En{\ss}lin}, T.~A. and {Eriksen}, H.~K. and {Fantaye}, Y. and {Fernandez-Cobos}, R. and {Finelli}, F. and {Forastieri}, F. and {Frailis}, M. and {Fraisse}, A.~A. and {Franceschi}, E. and {Frolov}, A. and {Galeotta}, S. and {Galli}, S. and {Ganga}, K. and {G{\'e}nova-Santos}, R.~T. and {Gerbino}, M. and {Ghosh}, T. and {Gonz{\'a}lez-Nuevo}, J. and {G{\'o}rski}, K.~M. and {Gratton}, S. and {Gruppuso}, A. and {Gudmundsson}, J.~E. and {Hamann}, J. and {Handley}, W. and {Hansen}, F.~K. and {Herranz}, D. and {Hivon}, E. and {Huang}, Z. and {Jaffe}, A.~H. and {Jones}, W.~C. and {Karakci}, A. and {Keih{\"a}nen}, E. and {Keskitalo}, R. and {Kiiveri}, K. and {Kim}, J. and {Knox}, L. and {Krachmalnicoff}, N. and {Kunz}, M. and {Kurki-Suonio}, H. and {Lagache}, G. and {Lamarre}, J.-M. and {Lasenby}, A. and {Lattanzi}, M. and {Lawrence}, C.~R. and {Le Jeune}, M. and {Levrier}, F. and {Lewis}, A. and {Liguori}, M. and {Lilje}, P.~B. and {Lindholm}, V. and {L{\'o}pez-Caniego}, M. and {Lubin}, P.~M. and {Ma}, Y.-Z. and {Mac{\'\i}as-P{\'e}rez}, J.~F. and {Maggio}, G. and {Maino}, D. and {Mandolesi}, N. and {Mangilli}, A. and {Marcos-Caballero}, A. and {Maris}, M. and {Martin}, P.~G. and {Mart{\'\i}nez-Gonz{\'a}lez}, E. and {Matarrese}, S. and {Mauri}, N. and {McEwen}, J.~D. and {Melchiorri}, A. and {Mennella}, A. and {Migliaccio}, M. and {Miville-Desch{\^e}nes}, M.-A. and {Molinari}, D. and {Moneti}, A. and {Montier}, L. and {Morgante}, G. and {Moss}, A. and {Natoli}, P. and {Pagano}, L. and {Paoletti}, D. and {Partridge}, B. and {Patanchon}, G. and {Perrotta}, F. and {Pettorino}, V. and {Piacentini}, F. and {Polastri}, L. and {Polenta}, G. and {Puget}, J.-L. and {Rachen}, J.~P. and {Reinecke}, M. and {Remazeilles}, M. and {Renzi}, A. and {Rocha}, G. and {Rosset}, C. and {Roudier}, G. and {Rubi{\~n}o-Mart{\'\i}n}, J.~A. and {Ruiz-Granados}, B. and {Salvati}, L. and {Sandri}, M. and {Savelainen}, M. and {Scott}, D. and {Sirignano}, C. and {Sunyaev}, R. and {Suur-Uski}, A.-S. and {Tauber}, J.~A. and {Tavagnacco}, D. and {Tenti}, M. and {Toffolatti}, L. and {Tomasi}, M. and {Trombetti}, T. and {Valiviita}, J. and {Van Tent}, B. and {Vielva}, P. and {Villa}, F. and {Vittorio}, N. and {Wandelt}, B.~D. and {Wehus}, I.~K. and {White}, M. and {White}, S.~D.~M. and {Zacchei}, A. and {Zonca}, A.},
        title = "{Planck 2018 results. VIII. Gravitational lensing}",
      journal = {\aap},
         year = 2020,
        month = sep,
       volume = {641},
          eid = {A8},
        pages = {A8},
          doi = {10.1051/0004-6361/201833886},
archivePrefix = {arXiv},
       eprint = {1807.06210},
 primaryClass = {astro-ph.CO},
       adsurl = {https://ui.adsabs.harvard.edu/abs/2020A&A...641A...8P}
}

@ARTICLE{2015MNRAS.447..499M,
       author = {{McGreer}, Ian D. and {Mesinger}, Andrei and {D'Odorico}, Valentina},
        title = "{Model-independent evidence in favour of an end to reionization by z {\ensuremath{\approx}} 6}",
      journal = {\mnras},
         year = 2015,
        month = feb,
       volume = {447},
       number = {1},
        pages = {499-505},
          doi = {10.1093/mnras/stu2449},
archivePrefix = {arXiv},
       eprint = {1411.5375},
 primaryClass = {astro-ph.CO},
       adsurl = {https://ui.adsabs.harvard.edu/abs/2015MNRAS.447..499M}
}

@ARTICLE{2025PhRvD.112j3048C,
       author = {{Cang}, Junsong and {Gao}, Yu and {Ma}, Yin-Zhe},
        title = "{Signatures of inhomogeneous dark matter annihilation on 21-cm}",
      journal = {\prd},
         year = 2025,
        month = nov,
       volume = {112},
       number = {10},
          eid = {103048},
        pages = {103048},
          doi = {10.1103/69jk-m5xg},
archivePrefix = {arXiv},
       eprint = {2312.17499},
 primaryClass = {astro-ph.CO},
       adsurl = {https://ui.adsabs.harvard.edu/abs/2025PhRvD.112j3048C}
}

@ARTICLE{2025JCAP...04..081H,
       author = {{Hou}, Liqiang and {Mack}, Katherine J.},
        title = "{The effects of dark matter annihilation and dark matter-baryon velocity offsets at Cosmic Dawn}",
      journal = {\jcap},
         year = 2025,
        month = apr,
       volume = {2025},
       number = {4},
          eid = {081},
        pages = {081},
          doi = {10.1088/1475-7516/2025/04/081},
archivePrefix = {arXiv},
       eprint = {2411.10626},
 primaryClass = {astro-ph.CO},
       adsurl = {https://ui.adsabs.harvard.edu/abs/2025JCAP...04..081H}
}

@ARTICLE{2024PhRvD.110l3529X,
       author = {{Xu}, Clara and {Qin}, Wenzer and {Slatyer}, Tracy R.},
        title = "{CMB limits on decaying dark matter beyond the ionization threshold}",
      journal = {\prd},
         year = 2024,
        month = dec,
       volume = {110},
       number = {12},
          eid = {123529},
        pages = {123529},
          doi = {10.1103/PhysRevD.110.123529},
archivePrefix = {arXiv},
       eprint = {2408.13305},
 primaryClass = {astro-ph.CO},
       adsurl = {https://ui.adsabs.harvard.edu/abs/2024PhRvD.110l3529X}
}

@ARTICLE{2026JCAP...02..060K,
       author = {{Koivu}, Emily and {Gnedin}, Nickolay Y. and {Hirata}, Christopher M.},
        title = "{Effects of primordial black holes on IGM history}",
      journal = {\jcap},
         year = 2026,
        month = feb,
       volume = {2026},
       number = {2},
          eid = {060},
        pages = {060},
          doi = {10.1088/1475-7516/2026/02/060},
archivePrefix = {arXiv},
       eprint = {2510.00246},
 primaryClass = {astro-ph.CO},
       adsurl = {https://ui.adsabs.harvard.edu/abs/2026JCAP...02..060K}
}

@ARTICLE{2026PhRvD.113f3031K,
       author = {{Klipfel}, Alexandra P. and {Kaiser}, David I.},
        title = "{Gravitational ionization by Schwarzschild primordial black holes}",
      journal = {\prd},
         year = 2026,
        month = mar,
       volume = {113},
       number = {6},
          eid = {063031},
        pages = {063031},
          doi = {10.1103/4p67-rxhp},
archivePrefix = {arXiv},
       eprint = {2601.05935},
 primaryClass = {hep-ph},
       adsurl = {https://ui.adsabs.harvard.edu/abs/2026PhRvD.113f3031K}
}

@ARTICLE{2021ApJ...919..120M,
       author = {{Morales}, Alexa M. and {Mason}, Charlotte A. and {Bruton}, Sean and {Gronke}, Max and {Haardt}, Francesco and {Scarlata}, Claudia},
        title = "{The Evolution of the Lyman-alpha Luminosity Function during Reionization}",
      journal = {\apj},
         year = 2021,
        month = oct,
       volume = {919},
       number = {2},
          eid = {120},
        pages = {120},
          doi = {10.3847/1538-4357/ac1104},
archivePrefix = {arXiv},
       eprint = {2101.01205},
 primaryClass = {astro-ph.GA},
       adsurl = {https://ui.adsabs.harvard.edu/abs/2021ApJ...919..120M}
}

@ARTICLE{2025ApJS..278...33K,
       author = {{Kageura}, Yuta and {Ouchi}, Masami and {Nakane}, Minami and {Umeda}, Hiroya and {Harikane}, Yuichi and {Yoshiura}, Shintaro and {Nakajima}, Kimihiko and {Yajima}, Hidenobu and {Thai}, Tran Thi},
        title = "{Census of Ly{\ensuremath{\alpha}} Emission from {\ensuremath{\sim}}600 Galaxies at z = 5-14: Evolution of the Ly{\ensuremath{\alpha}} Luminosity Function and a Late Sharp Cosmic Reionization}",
      journal = {\apjs},
         year = 2025,
        month = jun,
       volume = {278},
       number = {2},
          eid = {33},
        pages = {33},
          doi = {10.3847/1538-4365/adc690},
archivePrefix = {arXiv},
       eprint = {2501.05834},
 primaryClass = {astro-ph.GA},
       adsurl = {https://ui.adsabs.harvard.edu/abs/2025ApJS..278...33K}
}

@ARTICLE{2026A&A...705A.114M,
       author = {{Mason}, Charlotte A. and {Chen}, Zuyi and {Stark}, Daniel P. and {Yi Lu}, Ting and {Topping}, Michael and {Tang}, Mengtao},
        title = "{Constraints on the z {\ensuremath{\sim}} 6{\ensuremath{-}}13 intergalactic medium from JWST spectroscopy of Lyman-alpha damping wings in galaxies}",
      journal = {\aap},
         year = 2026,
        month = jan,
       volume = {705},
          eid = {A114},
        pages = {A114},
          doi = {10.1051/0004-6361/202553820},
archivePrefix = {arXiv},
       eprint = {2501.11702},
 primaryClass = {astro-ph.GA},
       adsurl = {https://ui.adsabs.harvard.edu/abs/2026A&A...705A.114M}
}

@ARTICLE{2026ApJ...997...86U,
       author = {{Umeda}, Hiroya and {Ouchi}, Masami and {Kageura}, Yuta and {Harikane}, Yuichi and {Nakane}, Minami and {Thai}, Tran Thi and {Nakajima}, Kimihiko},
        title = "{Probing the Cosmic Reionization History with JWST: Gunn-Peterson and Ly{\ensuremath{\alpha}} Damping Wing Absorption at 4.5 < z < 13}",
      journal = {\apj},
         year = 2026,
        month = jan,
       volume = {997},
       number = {1},
          eid = {86},
        pages = {86},
          doi = {10.3847/1538-4357/ae232b},
archivePrefix = {arXiv},
       eprint = {2504.04683},
 primaryClass = {astro-ph.GA},
       adsurl = {https://ui.adsabs.harvard.edu/abs/2026ApJ...997...86U}
}

@ARTICLE{2025JCAP...08..082A,
       author = {{Allali}, Itamar J. and {Singh}, Praniti and {Fan}, JiJi and {Li}, Lingfeng},
        title = "{Reionization and the Hubble constant: correlations in the Cosmic Microwave Background}",
      journal = {\jcap},
         year = 2025,
        month = aug,
       volume = {2025},
       number = {8},
          eid = {082},
        pages = {082},
          doi = {10.1088/1475-7516/2025/08/082},
archivePrefix = {arXiv},
       eprint = {2503.05691},
 primaryClass = {astro-ph.CO},
       adsurl = {https://ui.adsabs.harvard.edu/abs/2025JCAP...08..082A}
}

@ARTICLE{2026PhRvD.113h3515A,
       author = {{Allali}, Itamar J. and {Li}, Lingfeng and {Singh}, Praniti and {Fan}, JiJi},
        title = "{Cosmic tensions indirectly correlate with reionization optical depth}",
      journal = {\prd},
         year = 2026,
        month = apr,
       volume = {113},
       number = {8},
          eid = {083515},
        pages = {083515},
          doi = {10.1103/14tf-4nk2},
archivePrefix = {arXiv},
       eprint = {2509.09678},
 primaryClass = {astro-ph.CO},
       adsurl = {https://ui.adsabs.harvard.edu/abs/2026PhRvD.113h3515A}
}

@ARTICLE{2024JHEP...12..216E,
       author = {{Esteban}, Ivan and {Gonzalez-Garcia}, M.~C. and {Maltoni}, Michele and {Martinez-Soler}, Ivan and {Pinheiro}, Jo{\~a}o Paulo and {Schwetz}, Thomas},
        title = "{NuFit-6.0: updated global analysis of three-flavor neutrino oscillations}",
      journal = {Journal of High Energy Physics},
         year = 2024,
        month = dec,
       volume = {2024},
       number = {12},
          eid = {216},
        pages = {216},
          doi = {10.1007/JHEP12(2024)216},
archivePrefix = {arXiv},
       eprint = {2410.05380},
 primaryClass = {hep-ph},
       adsurl = {https://ui.adsabs.harvard.edu/abs/2024JHEP...12..216E}
}

@article{Gompertz1825,
  title={XXIV. On the nature of the function expressive of the law of human mortality, and on a new mode of determining the value of life contingencies. In a letter to Francis Baily, Esq. FRS \&c},
  author={Gompertz, Benjamin},
  journal={Philosophical transactions of the Royal Society of London},
  number={115},
  pages={513--583},
  year={1825},
  publisher={The Royal Society London}
}

@ARTICLE{2024ApJ...976L..11R,
       author = {{Roy Choudhury}, Shouvik and {Okumura}, Teppei},
        title = "{Updated Cosmological Constraints in Extended Parameter Space with Planck PR4, DESI Baryon Acoustic Oscillations, and Supernovae: Dynamical Dark Energy, Neutrino Masses, Lensing Anomaly, and the Hubble Tension}",
      journal = {\apjl},
         year = 2024,
        month = nov,
       volume = {976},
       number = {1},
          eid = {L11},
        pages = {L11},
          doi = {10.3847/2041-8213/ad8c26},
archivePrefix = {arXiv},
       eprint = {2409.13022},
 primaryClass = {astro-ph.CO},
       adsurl = {https://ui.adsabs.harvard.edu/abs/2024ApJ...976L..11R}
}

@ARTICLE{2025ApJ...986L..31R,
       author = {{Roy Choudhury}, Shouvik},
        title = "{Cosmology in Extended Parameter Space with DESI Data Release 2 Baryon Acoustic Oscillations: A 2{\ensuremath{\sigma}}+ Detection of Nonzero Neutrino Masses with an Update on Dynamical Dark Energy and Lensing Anomaly}",
      journal = {\apjl},
         year = 2025,
        month = jun,
       volume = {986},
       number = {2},
          eid = {L31},
        pages = {L31},
          doi = {10.3847/2041-8213/ade1cc},
archivePrefix = {arXiv},
       eprint = {2504.15340},
 primaryClass = {astro-ph.CO},
       adsurl = {https://ui.adsabs.harvard.edu/abs/2025ApJ...986L..31R}
}

@ARTICLE{2025ApJ...994L..26R,
       author = {{Roy Choudhury}, Shouvik and {Okumura}, Teppei and {Umetsu}, Keiichi},
        title = "{Cosmological Constraints on Nonphantom Dynamical Dark Energy with DESI Data Release 2 Baryon Acoustic Oscillations: A 3{\ensuremath{\sigma}}+ Lensing Anomaly}",
      journal = {\apjl},
         year = 2025,
        month = nov,
       volume = {994},
       number = {1},
          eid = {L26},
        pages = {L26},
          doi = {10.3847/2041-8213/ae1a64},
archivePrefix = {arXiv},
       eprint = {2509.26144},
 primaryClass = {astro-ph.CO},
       adsurl = {https://ui.adsabs.harvard.edu/abs/2025ApJ...994L..26R}
}
\bibliographystyle{scpma_short}

\appendix
\section{Gompertzian reionization}
\label{app:gomp}
Here we provide a brief description of the Gompertzian reionization model for the purpose of self-consistency. For details and discovery, see \cite{2024arXiv240513680M}. 

The Gompertzian curve is an asymmetric sigmoid function originally introduced to analyze age-dependent mortality \cite{Gompertz1825}. The Gompertzian reionization model easily matches EoR constraints and has the expected slow start with a following rapid acceleration past its midpoint. 

The expression for the neutral hydrogen evolution is given by
\begin{align}
    &x_\HI(\tilde{a}) = \exp{\{-\exp[P_5(\tilde{a})]\}} \, , \label{eq:gomp} \\
    &P_5(\tilde{a}) = \sum_{m = 0}^{5} c_m \ln^m \tilde{a} \, , \\
    &\boldsymbol{c} = \{0, 1, 0.1252, 0.03533, 0.002203, 0.000007483 \}\, , \\
    &\tilde{a}(a) = \left[\frac{a}{\alpha} \right]^{\beta}\, \label{eq:a_rescale} ,
\end{align}
where $a$ is the scale factor, $\alpha$ is the power-law pivot, and $\beta$ a rescaling tilt. Furthermore, the 5th-degree polynomial is introduced to account for residuals from the Gompertz shape and its coefficients were calibrated using simulations \cite{2026arXiv260413423M}.

\end{multicols}
\end{document}